\documentstyle[twoside,fleqn,espcrc2]{article}

\def\gtap{\;\raisebox{-.5ex}{\rlap{$\sim$}} \raisebox{.5ex}{$>$}\;}

\begin{document}



\title{QCD Sum Rules on the Lattice}

\author{Chris R. Allton\address{
Department of Physics, University of Wales Swansea, U.K. \\
$^{\rm b}$ DESY - Theory Group, Hamburg, Germany\\}
\thanks{Talk presented by C.R. Allton, Preprint SWAT/163}
and Stefano Capitani$^{\rm b}$}

\begin{abstract}
We study the work of Leinweber by applying the Continuum
Model of QCD Sum Rules (QCDSR) to the analysis of (quenched) lattice
correlation functions. We expand upon his work in several areas
and find that, while the QCDSR Continuum Model very adequately
fits lattice data, it does so only for non-physical values of
its parameters.
The non-relativistic model is found to predict
essentially the same form for the correlation functions as the QCDSR
Continuum Model but without the latter's restrictions.
By fitting lattice data to a general form which includes
the non-relativistic quark model as a special case, we confirm
it as the model of choice.
\end{abstract}

\maketitle



\section{INTRODUCTION}

Lattice Gauge Theory and QCD Sum Rules (QCDSR) are two areas of
research which have been widely employed to deepen the understanding of
systems governed by the strong interactions. In this paper we combine
both methods following the pioneering work of Leinweber \cite{lw1,lw2}.
Utilising known results in QCD Sum Rules we apply them to Lattice Gauge
Theory in order to check that the two are compatible.

We extend the work of \cite{lw1,lw2} by including mesonic states,
by applying our analysis to several $\beta$ values,
and by studying data from both the Wilson and SW-Clover actions.
We also study the non-relativistic quark model predictions.

The lattice data used is from the APE collaboration and is a mixture
of (quenched) Wilson and SW-Clover with beta values ranging from
$\beta=6.0$ to $\beta=6.4$.  Zero-momentum 2-point hadronic correlation
functions were studied for the nucleon, delta and the following
mesonic operators: $P_5$, $A_0$ \& $V_i$ (in an obvious notation).  A
longer version of this work appears in \cite{long}.


\section{QCDSR CONTINUUM MODEL}
\label{cont_model}

The quark propagator in Euclidean space's Wilson OPE reads
(in the coordinate gauge),
\begin{equation}
S^{aa'}_q=\{
           \frac{\gamma \cdot x}{2 \pi^2 x^4} +
           \frac{m_q}{(2 \pi x)^2}            -
           \frac{\langle : \overline{q}q : \rangle}{2^2 3}
          \}  \delta^{aa'} +
          \cdots.
\label{eq:qprop}
\end{equation}
This is substituted into the (Wick contracted) 2-point function
leading to the generic form:
\begin{equation}
G_2^{OPE}(t)= \sum_{n=-\infty}^{n_0} \frac{1}{t^n} \cdot C_n
O_n  (m_q, \langle : \overline{q}q : \rangle )
\label{eq:expans},
\end{equation}
where $C_n$ is a numerical coefficient, $O_n$ is some operator function,
and $n_0$ is a positive integer.
It is now useful to express the timesliced correlation function,
$G_2(t)$, in the spectral representation:
$G_2(t)=\int_0^{\infty} \rho (s) e^{-st} ds.$
The OPE spectral density, $\rho^{OPE}(s)$, is
calculated by means of a simple inverse Laplace transform of
$G_2^{OPE}$,
\begin{equation}
\rho^{OPE}(s) = \sum_{n=1}^{n_0} \frac{s^{n-1}}{(n-1)!} \cdot C_n
O_n  (m_q, \langle : \overline{q}q : \rangle ),
\label{eq:rho}
\end{equation}
where $Re(t) > 0$. In the sum over $n$, we have kept only the positive
values of $n$ since we are interested in the leading terms as $t
\rightarrow 0$.

The QCD Continuum Model is introduced by setting a
threshold $s_0$ in the energy scale $s$,
so that the excited states' contribution to
$G_2(t)$ is given only by the energies above that scale. Performing the
integral over $s$ we obtain,
\begin{eqnarray} \nonumber
G_2^{cont}(t) &\equiv& \int_{s_0}^{\infty} \rho^{OPE}(s) e^{-st} ds \\
&=& e^{-s_0t} \sum_{n=1}^{n_0}
\sum_{k=0}^{n-1} \frac{1}{k!}\frac{s_0^k}{t^{n-k}}
\cdot C_n O_n.
\label{eq:g2_cont}
\end{eqnarray}
The full correlation function contains the ground state as well as the
above continuum contribution $G_2^{cont}$.
This is simply included as a $\delta$ function in $\rho(s)$,
leading to the full correlation function
\begin{equation}
G_2^{\mbox{Full}} (t) = \frac{Z}{2M} e^{-Mt} + \xi
\int_{s_0}^{\infty} \rho^{OPE}(s) e^{-st} ds.
\label{eq:model}
\end{equation}
There are four parameters in this Ansatz for $G_2^{\mbox{Full}}(t)$:
$Z$ and $M$ for the ground state, $s_0$ (the continuum threshold), and
$\xi$.  The parameter $\xi$ is introduced to allow for lattice
``distortions'' in the normalisation of the excited states
\cite{lw1}. In the continuum limit $\xi$ should be equal to one.

We have calculated the OPE expansions for all the channels studied.
These are listed in \cite{long}.

\section{FITS TO DATA}
\label{qsr_fits}


Correlation functions were fitted to the following three
functional forms, $F(t)$:

\noindent{\bf 1. Conventional Single State Fit {\em (``1-exp'')}}
i.e. using $F(t) = \frac{Z}{2M} e^{-Mt}$.

\noindent{\bf 2. QCD Continuum Model Fit {\em (``Cont'')}}
These are described in the previous section.

\noindent{\bf 3. Conventional Two State Fit {\em (``2-exp'')}}
i.e. using $F(t) = \frac{Z}{2M} e^{-Mt} + \frac{Z'}{2M'} e^{-M't}$.
This functional form was chosen since it is traditionally used as the
generalisation of ``1-exp'' to include the higher mass state(s).
It has the same number of parameters as the ``Cont'' fit
and the results of these two fits will be directly compared.

The ``1-exp'' fits are used here to gain ``standard values'' for
$Z$ and $M$ and are fitted in the asymptotic region only.
In the case of the QCD Continuum Model fits, we use time windows which
begin very close to the origin, e.g. $t=2-28$ for $\beta=6.0$.  So
that a direct comparison can be performed, the same time windows are
used to perform the ``2-exp'' fits.

The full results of the fits to the three fitting functions for all
the lattice correlation data are presented in \cite{long}.
We show in Table~\ref{tab:vi}, the fits for the vector meson
channel for the Wilson $\beta=6.1$ only.
(The other fits follow a similar pattern.)

The main features of the fits are now summarised. The Continuum Model
reproduces the data better than the 2-exp case, i.e. (i) its $\chi^2$
values are lower (sometimes by an order of magnitude), and,
(ii) its ground state parameters $Z$ and $M$ are closer to the ``1-exp''
control case.

We also note that the 2-exp fit values over-estimate $Z$ and $M$ in
every channel studied.  In \cite{long} we give an explanation for this
effect.

Clearly it is not sufficient for the ``Cont'' fits to reproduce the
ground state parameters $Z$ and $M$, and to have a sensible $\chi^2$,
they must also give values for the continuum parameters, $s_0$ and
$\xi$ which are acceptable within the assumptions of the QCDSR
Continuum Model. These are: (a) since $s_0$ corresponds to a physical
threshold, they should be constant (in $GeV$) for each channel as
$\beta$ is varied; (b) $s_0$ should be large enough to be in a region
where perturbation theory is valid; and (c) $\xi$ should scale like
$\xi \rightarrow 1$ as $a \rightarrow 0$.
It turns out that criteria (a) is satisfied for mesons, but not for baryons;
that (b) is marginal; and that (c) fails totally \cite{long}.
(In fact $\xi$ can be large and negative !)

This leads us to the conclusion that the applicability of the QCDSR
Continuum Model to lattice data is questionable since it leads to values
for $s_0$ and $\xi$ which are inconsistent with the assumptions made.

It is clear that, apart from taking on non-physical values of
its fitting parameters, the {\em ``Cont''} Ansatz does reproduce the
lattice correlation function data. Therefore we require similar
fitting Ans\"atze as those in the {\em ``Cont''} case, but without the
corresponding restrictions on the parameters $s_0$ and $\xi$.



\begin{table}
\begin{tabular}{lccc}
\hline
$F(t)$			& 1-exp		& Cont		& 2-exp		\\
\hline
$Z (\times 10^{-2})$	& 0.57(3)	& 0.60(2)	& 0.68(2)	\\
$Ma$			& 0.466(2)	& 0.468(2)	& 0.474(2)	\\
$s_0$ [GeV]		& 		& 1.9(1)	&		\\
$\xi$			&		& -3.25(4)	&		\\
$\chi^2/d.o.f.$		& 0.4(6)/11	& 9(5)/22	& 110(20)/22	\\
\hline
\end{tabular}
\caption{ \it{Values for the fitting parameters for the vector meson
(Wilson $\beta=6.1$ case).}}
\label{tab:vi}
\end{table}



\section{QUARK MODELS}
\label{qm_fits}

We present the derivation of the density of states, $\rho(s)$, for
hadronic correlators for the non-relativistic quark model.  Taking the
quarks as on-shell, there are 3 degrees of freedom available for each
quark in the hadron: one per spatial momentum component.  Therefore in
the case of mesons there are naively 6 degrees of freedom, with 9
degrees of freedom for baryons. However, there are 4 constraint
equations corresponding to the condition that the hadron's 4-momentum is
zero, leaving 2 degrees of freedom for the mesons and 5 degrees
of freedom for the baryons. This simple, non-relativistic analysis
leads to $\rho_{nrqm}(s) \sim s^2$ for mesons, and
$\rho_{nrqm}(s) \sim s^5$ for baryons.
This continuum-like behaviour for $\rho(s)$ is clearly most
appropriate for large $s$ where the quarks are approximately free.
Again we approximate the actual density of states by
$\rho(s) = \frac{Z}{2M} \; \delta(s-M) +
 \theta(s-s_0) \rho_{nrqm}(s)$,
where the delta-function represents the ground state and the non-relativistic
quark model result is used for the continuum (beginning at the threshold
energy $s_0$).

Using this definition of
$\rho(s)$ in the spectral equation we obtain,
for mesons, 
\[
G_2(t) = \frac{Z}{2M} e^{-Mt} + K \left(
\frac{1}{t^{3}} + \frac{s_0}{t^2} + \frac{s_0^2}{2t}
\right) e^{-s_0 t},
\]
where $K$ is some numerical constant,
with a similar expression for baryons.
It is very interesting to note that these expressions are {\em identical}
to those obtained for the Cont model with $m_q$ and the condensate set to zero.
Since the terms ${\cal O}(m_q,\langle : \overline{q}q : \rangle)$
are numerically insignificant, we have shown that the non-relativistic
quark model predicts essentially the same functional form as the QCDSR's
Continuum Model. Thus we have achieved the aim raised in the previous Section of
a (physically motivated) functional form which reproduces the data
better than the ``2-exp'' fits, but which doesn't
suffer from unphysical parameter values such as the ``Cont'' fit.
(The relativistic quark model is discussed at length in \cite{long}.)

It is natural to wonder if there is any way of using the lattice
correlation function data to pin down the best form of the density of
states $\rho(s)$. With this in mind we assume that the density of states
has the following form:
$\rho(s) = \frac{Z}{2M} \delta(s-M) + \theta(s-s_0) K s^n$,
where $n$ is to be determined from the fit. The two-point correlation function
for this Ansatz is defined from the standard Laplace transform.
Studying the $\chi^2$ values as a function of $n$ we see
a clear
minimum in the case of the pseudoscalar mesons at $n=2$.
The vector channel's minimum is more spread
out at around $1\le n \le 3$.
The proton case is less clear again; it has a minimum at
$n \gtap 5$.
These results confirm the predictions of the non-relativistic quark model.

\section{DISCUSSION \& CONCLUSION}

This work began with a study of the QCD Sum Rule Continuum Model as
applied to lattice two-point correlation functions using the method
first introduced by Leinweber \cite{lw1,lw2}.  We have extended his work
by including mesonic states, by fitting lattice data at several
lattice spacings and by using two formulations of the lattice
action. We have found that the QCD Sum Rule Continuum Model
successfully fits the lattice data, but only for unphysical values of
its fitting parameters. This leads us to the conclusion (somewhat
contrary to that of \cite{lw1}) that this model cannot
self-consistently fit lattice data.

We have then searched for a model that reproduces a similar functional form
as the QCD Continuum Model, but which does not have its restrictions on its
parameters. The non-relativistic quark model was studied and found
to be such a theory.
We introduced a general fitting function which has the non-relativistic
quark model as a special case, and found evidence that the $\chi^2$ was
minimized for the fitting function corresponding roughly with the
non-relativistic quark model.

We propose that these fitting functions can be used to extract ground
state properties in cases where the asymptotic state is heavily contaminated
by excited states, or where the time separation in the
correlators is forced to be small.


\section{ACKNOWLEDGEMENTS}
This work was supported by a grant form the Nuffield Foundation.





\begin{thebibliography}{99}

\bibitem{lw1} D.B.Leinweber, Phys. Rev. {\bf D51} (1995) 6369

\bibitem{lw2} D.B.Leinweber, Phys. Rev. {\bf D51} (1995) 6383

\bibitem{long} C.R. Allton and S. Capitani {\em in preparation}

\end{thebibliography}
\end{document}